\documentclass[prl,twocolumn,showpacs,preprintnumbers,amsmath,amssymb]{revtex4}

\usepackage{graphicx}
\usepackage{dcolumn}
\usepackage{bm}

\newcommand{\ket}[1]{|{#1}\rangle}
\newcommand{\bra}[1]{\langle{#1}|}

\begin{document}

\preprint{APS/123-QED}

\title{Opacity of electromagnetically induced transparency for quantum fluctuations}

\author{P. Barberis-Blostein}
 \email{pbb@fis.unam.mx}
\author{M. Bienert}%
 \email{bienert@fis.unam.mx}
\affiliation{
Centro de Ciencias F{\'\i}sicas, Universidad Nacional Autonoma de
M\'exico, Campus Morelos UNAM, 62251 Cuernavaca, Morelos, M\'exico
}

\date{\today}

\begin{abstract}
We analyze the propagation of a pair of quantized fields inside a medium of three-level
atoms in $\Lambda$ configuration. We calculate the stationary
quadrature noise spectrum of the field after propagating through the medium, in
the case where the probe field is in a squeezed state and the atoms show
electromagnetically induced transparency (EIT). We find an oscillatory
transfer of the initial quantum properties between the probe and pump
fields which is most strongly pronounced when both fields have comparable Rabi
frequencies. This implies that the quantum state measured after propagation can
be completely different from the initial state, even though the mean values of the
field are unaltered.
\end{abstract}

                            

\maketitle

\begin{figure}
\includegraphics[width=5cm]{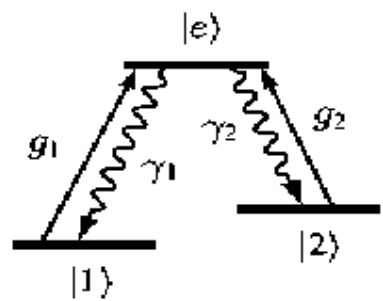}
\caption{\label{fig:model} The atoms have a $\Lambda$
  configuration with stable or metastable states $\ket 1$ and $\ket 2$ and
  common excited state $\ket e$. The transitions $\ket i\leftrightarrow \ket
  e$ with dipole coupling constants $g_j$ underlie spontaneous decay of rates
  $\gamma_j$ $(j=1,2)$, the linewidth of $\ket e$ is
  $\gamma=\gamma_1+\gamma_2$.}
\end{figure}

\begin{figure}
\includegraphics[width=7.5cm]{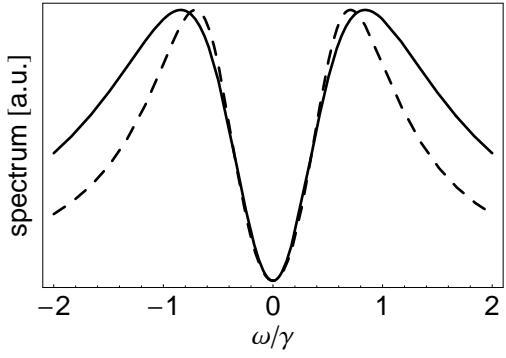}
\caption{\label{fig:res} Absorption spectrum of the probe's mean value (solid
  line) and the function $P(\omega,0)$, Eq.~\eqref{eq:p}, occurring in the
  spectrum of quantum fluctuations (dashed line), in arbitrary units. Here,
  $\omega=0$ refers to a resonant driving of the probe transition.
 Both curves are normalized to have the same maximal values. Their positions
$\omega^{(\!-\!\!-\!\!-)}_{\rm
  max}=\frac{(\Omega_1^2+\Omega_2^2)^{3/4}}{\sqrt{\Omega_1}}$ and $\omega_{\rm
  max}^{(\text{- - -})}=\sqrt{\Omega^2_1+\Omega^2_2}$ are slightly
  shifted. (Parameters: $\Omega_1=\Omega_2=\gamma, g_1=g_2.$)}
\end{figure}

Electromagnetically induced transparency (EIT) \cite{rv:harris} is a technique
that can be used to eliminate fluorescence from an atom
illuminated with light whose frequency is equal to a particular atomic
transition. This technique can be used in systems of
three-level atoms in $\Lambda$ configuration \cite{rv:marangos}, see Fig.~\ref{fig:model}. In
this configuration a mode of the field, called the pump field, interacts
resonantly with one dipole transition, while another mode, the probe field,
interacting with the second dipole transition, is tested for transparency. The
linear response of the absorption of the probe field by the medium is
described by the imaginary part of the electric susceptibility. In
Fig.~\ref{fig:res} we plot the susceptibility as a function of
the probe frequency (solid line). The maximum absorption of the probe field by the medium
depends on the Rabi frequencies associated with each atomic optical
transition. The maximum occurs for a detuning from resonance which increases
monotonically with the Rabi frequencies.

Many recent works have investigated if this transparency,
originally studied for classical fields, preserves the initial quantum
properties of the probe field. For a classical pump field with Rabi frequency
much larger than that of 
the probe, Lukin {\it et al.}~\cite{rv:memoria2} showed that the
medium is transparent for the quantum state. Furthermore, they demonstrated
a transfer of the initial quantum state from the probe field
to the atoms and in a second stage back to the field by varying the Rabi
frequency of the pump laser. They proposed using this technique
as a quantum memory device. When both fields are treated quantum
mechanically, Dantan {\it et al.}~\cite{rv:dantan4} studied the noise
spectrum of the quadratures, when only the coherent pump field drives the
atoms and the
probe field is initially in a broad-band squeezed vacuum. If the frequency equals
that of the atomic transition, the  medium is
transparent. For other frequencies, there is absorption of the quantum
properties. The spectral absorption
varies in a similar manner as the transparency curve for
the mean value of the field.
The transparency for the vacuum squeezed state was confirmed experimentally
by Akamatsu {\it et al.}~\cite{rv:kozuma}. 

In this Letter, we discuss the propagation of a quantum state in a medium of
  three level atoms showing EIT in the stationary regime. We focus on the case
  of a squeezed initial state of the probe field and a coherent pump field. We
  treat both fields quantum mechanically and in contrast to previous work we do not put any constraint on the probe Rabi frequency. An analytical solution is given for this general case. Our main result is
  that the absorption of the initial squeezing is accompanied by an
  oscillatory interchange of the quantum properties between the pump and probe
  field while traveling through the medium. This oscillatory behavior is
  present for frequencies where the absorption of the mean value of
  the field is usually negligible. The effect is maximally pronounced for the case of equal Rabi frequencies. As a consequence, the probe and pump states
  after propagation can be {\it completely different} from the input
  state. 

To probe the behavior of quantum fluctuations of light passing through an EIT--medium, we use squeezed states defined by
$\ket{\alpha;\xi}\equiv D(\alpha)S(\xi)\ket{0}$ 
with the squeezing operator $S(\xi) = \exp[(\xi {a^\dagger}^2-\xi^\ast
a^2)/2]$ and displacement operator $D(\alpha)=\exp[\alpha a^\dagger
-\alpha^\ast a]$, where $a$ and $a^\dagger$ are the annihilation and creation
operator of the mode under consideration. 
After propagation along the $z$--axis, we analyze the fluctuation spectrum
\begin{equation}
{\mathcal S}(\omega) = \int\limits_{-\infty}^\infty e^{-i\omega t} \langle \delta Y^\theta(t)\delta Y^\theta(0)\rangle
\label{eq:spec}
\end{equation}
in the steady state of the quadrature fluctuation
$\delta Y^\theta(t)=\delta a(t) \exp(-i\theta) + \delta a^\dagger(t) \exp(i\theta)$,
where $\delta a=a-\langle a\rangle$. 

We consider two quasi monochromatic one dimensional beams of light propagating
along the $z$ axis in a medium
of $N$ three-level atoms in $\Lambda$-configuration. The excited state $\ket e$ of the atom with total
linewidth $\gamma=\gamma_1+\gamma_2$ can decay spontaneously with rate
$\gamma_j$ into the lower electronic state $\ket j$ ($j=1,2$). To describe the
propagation of the two beams, we use a multimode
representation of a pair of electromagnetic fields $\vec E_j=\vec{\mathcal
  E}_j a_j(z,t) \exp[ik_{{\rm L},j}z-\omega_{{\rm L},j} t] + {\rm h.c.}$ and
treat the medium in a continuum approximation
\cite{bib:fleisch95,rv:dantan4}. Here, $|\vec{\mathcal E}_j|$ is the vacuum
electric field at the laser's carrier frequency $\omega_{{\rm L},j}=c k_{{\rm
    L},j}$, and $a_j(z,t)$ is the envelope operator of the corresponding field
$j$, which is slowly varying in space and time.
For the medium we assume the inter-atomic distance to be much smaller than the
shortest relevant wave length of the laser light and introduce the continuous atomic operators 
$\sigma_{\mu\nu}(z) = \lim_{\Delta z\rightarrow 0} \frac{L}{N\Delta z}\!\sum\limits_{z^{(j)}\in \Delta z }\! \sigma_{\mu\nu}^{(j)}$,
where $\sigma^{(j)}_{\mu\nu}=\ket{\mu}^{(j)}\bra{\nu}$ is the individual atomic operator of atom $j$ at position $z_j$.

The atoms interact with the electric fields via their dipole moment $\wp$; we
treat this interaction in the rotating wave approximation. Then, in the slowly
varying envelope approximation, the Heisenberg equation of motion for the field operator $a(z,t)$ gives
\begin{alignat}{1}
\left(\frac{\partial}{\partial t}+c\frac{\partial}{\partial z}\right)a_j = -i
g_j N \sigma_{je}\, ,
\stepcounter{equation}
\tag{\theequation a}
\label{eq:eprop}
\end{alignat}
where we have introduced the coupling constant $g_j=\vec \wp\vec{\mathcal
  E}_j/\hbar$. Using the inversion operators
$\varpi_j=\sigma_{ee}-\sigma_{jj}$, the Heisenberg-Langevin equations of the atomic degree of freedom for resonant driving read
\begin{alignat*}{1}
\frac{\partial}{\partial t} \varpi_1 =&\frac{1}{3}(-\gamma_1-\gamma)(1+ \varpi_{1}+\varpi_2)-2 i g_1(\sigma_{e1}a_1  - a^\dagger_1\sigma_{1e})\\ & - i g_2(\sigma_{e2}a_2-a^\dagger_2\sigma_{2e})  + f_{\varpi_1},\\
\frac{\partial}{\partial t}\varpi_2 =&\frac{1}{3}(-\gamma_2- \gamma)(1+\varpi_{1}+\varpi_2)
-i g_1(\sigma_{e1}a_1  -a^\dagger_1\sigma_{1e})\\ & - 2 i g_2 (\sigma_{e2}a_2-a^\dagger_2\sigma_{2e}) +f_{\varpi_2},\\
\frac{\partial}{\partial t} \sigma_{1e}=&-\frac{\gamma}{2}
\sigma_{1e}+i g_1 \varpi_1 a_1-i g_2 \sigma_{12} a_2
+f_{1e},\\
\frac{\partial}{\partial t} \sigma_{2e}=&-\frac{\gamma}{2}
\sigma_{2e}+i g_2 \varpi_2 a_2   -i g_1 \sigma_{21} a_1 
+ f_{2e} ,\\
\frac{\partial}{\partial t}\sigma_{21}=&-i g_1 a_1^\dagger \sigma_{2e} +i
g_2 \sigma_{e1} a_2. 
\tag{\theequation b}
\label{eq:aprop}
\end{alignat*}
This system of equations \eqref{eq:eprop} and \eqref{eq:aprop} is easily
interpreted: The polarization field $\sigma_{je}(z,t)$ serves as a source for
the electric fields, whereas the propagating light in turn drives the atomic
media via the dipole interaction terms $\propto g_j$. 
The $f_j$ are delta-correlated, collective Langevin operators which account for the noise
introduced by the coupling of the atomic system to the free radiation
field. They have vanishing mean values and correlation functions of the
form $\langle f_{x}(z,t)
f_{y}(z,t')\rangle=\frac{L}{N}D_{xy}\delta(t-t')\delta(z-z')$. The diffusion
coefficients $D_{xy}$ can be obtained from the generalized Einstein
equations \cite{bib:louisell}. For this particular problem they are listed in \cite{bib:pablonicim}

We will use the common technique \cite{bib:davidov,bib:louisell} of
transforming Eqs.~\eqref{eq:eprop} and \eqref{eq:aprop} into stochastic
c-number equations which serve to calculate correlation functions of the
atomic and field operator up to second order. To fix the corresponding order
of operators we use the ``normal'' order convention $ a_{2}^{\dagger },
a_{1}^{\dagger },
{\sigma}_{e2},{\sigma}_{e1},{\sigma}_{12},\varpi_{1},\varpi_{2},\sigma_{21},\sigma_{1e},
\sigma_{2e}, a_1, a_2$. All calculated results from the
c-number equations have to be considered to represent the results using
operators in this order. The equations of motion for the c-number
quantities are equivalent to those for the operators, except that a
redefinition of the diffusion coefficients is necessary. These new diffusion
coefficients are listed in \cite{bib:pablonicim}. In the
following, we use the same symbols for operators and their c-number
substitutions, except that for the field operators $a_j$ we instead use $\alpha_j$.

In order to solve the c-number counterparts of Eqs.~\eqref{eq:eprop} and
\eqref{eq:aprop}, it is convenient to collect the system quantities and the fluctuation forces in vectors
$
\mathbf{x}^{\rm T}(z,t) = (\alpha_{2}^{\ast }, \alpha_{1}^{\ast }, {\sigma}_{e2},\dots ,
\sigma_{2e}, \alpha_1, \alpha_2)$ and $
\mathbf{f}^{\rm T}(z,t) = (0,0,f_{e2}, \dots, f_{2e}, 0,0)
$,
where for the components we choose the same order convention introduced
above. For a large number of atoms, it is reasonable to assume a steady state for
the mean values $\langle \mathbf{x}\rangle$ with small fluctuations
$\delta\mathbf{x}$, i.e.~we write
$\mathbf{x}=\langle \mathbf{x}\rangle + \delta\mathbf{x}$. This allows us to
treat the problem perturbatively for small fluctuations
$\delta\mathbf{x}\propto O(1/\sqrt{N})$ \cite{bib:davidov,bib:stochasticmethods}. The steady state values
$\langle\mathbf{x}\rangle\propto O(1)$ are found from the equations of motion
\eqref{eq:eprop} and \eqref{eq:aprop} to zeroth order in $\delta\mathbf{x}$,
after setting all time derivatives to zero. Due to EIT the steady state values are $z$-independent. To first order, the system of equations 
\begin{equation}
\frac{\partial}{\partial t} \delta\mathbf{x}(s,t) = {\mathcal M}(s) \delta\mathbf{x}(s,t) + \mathbf{f}(s,t) + \mathbf{g}(t)
\label{eq:lineq}
\end{equation}
is linear in the fluctuations, where
$\delta\mathbf{x}(s,t)=\int\limits_0^\infty \exp[-sz] \delta\mathbf{x}(z,t)
dz$ (and analogously for $\mathbf{f}$) denotes the Laplace transform of
$\delta\mathbf{x}(z,t)$. The elements of the matrix ${\mathcal M}(s)$ are
calculated from the equations~\eqref{eq:eprop} and \eqref{eq:aprop}
to first order in $\delta \mathbf{x}$. In $\mathbf{g}(t)$ we collect the
initial conditions at $z=0$ following from the Laplace transform of
Eqs.~\eqref{eq:eprop}. Equation~\eqref{eq:lineq} can be cast into algebraic
form using the Fourier transform
$\delta\mathbf{x}(s,\omega)=\int\limits_{-\infty}^\infty dt\, \exp[-i\omega
t]\delta\mathbf{x}(s,\omega) /\sqrt{2\pi}$. From the solution
$\delta\mathbf{x}(s,\omega)=[{\mathcal M}+i\omega]^{-1}(\mathbf{f}+\mathbf{g})$ we can construct the correlation matrix 
\begin{alignat}{1}
\langle \delta\mathbf{x}(s,\omega)\delta\mathbf{x}^\dagger(s',\omega')\rangle=&\delta(\omega+\omega')\times\nonumber\\
[{\mathcal M(s)}+i\omega]^{-1}&\left({\mathcal D(s,s')+{\mathcal G}(\omega)}\right)[{\mathcal M^\dagger(s')}-i\omega]^{-1}
\label{eq:cmatrix}
\end{alignat}
with ${\mathcal
  D}\delta(\omega+\omega')=\langle\mathbf{f}\mathbf{f^\dagger}\rangle$,
${\mathcal
  G}\delta(\omega+\omega')=\langle\mathbf{g}\mathbf{g^\dagger}\rangle$ and
  $\langle \mathbf{f}\mathbf{g^\dagger}\rangle=\langle
  \mathbf{g}\mathbf{f^\dagger}\rangle=0$. For fields entering the medium in a
  squeezed state with real squeezing parameter $\xi_j$ ($j=1,2$), we have 
\begin{alignat}{1}
{\mathcal G}_{{\alpha_i},{\alpha_j}}={\mathcal
  G}_{{\alpha^\ast_i},{\alpha^\ast_j}}&=-c^2\delta_{ij}\cosh\xi_i\sinh\xi_i\, ,\\
{\mathcal G}_{{\alpha^\ast_i},{\alpha_j}}={\mathcal
  G}_{{\alpha_i},{\alpha^\ast_j}}&=c^2\delta_{i,j}\sinh^2\xi_i\, ,
\end{alignat}
with all other coefficients of ${\mathcal G}$ vanishing. 

We can now use the result \eqref{eq:cmatrix} to calculate the spectrum of fluctuations of the electric fields. To this end we recall the relation
\begin{equation}
\langle \delta Y_j^\theta(z,\omega) \delta
Y_j^\theta(z,\omega')\rangle=\delta(\omega+\omega') [{\mathcal
  S}_j(\omega)-1]\, ,
\label{eq:wk}
\end{equation}
following from the Wiener Khinchine theorem, connecting the fluctuation
spectrum $S_j(\omega)$, Eq.~\eqref{eq:spec}, with the correlation of the
quadrature fluctuations  $\delta Y_j^\theta(z,\omega)=\delta
\alpha_j(z,\omega) e^{-i\theta}+\delta
\alpha_j^\ast(z,-\omega)e^{i\theta}$. Recall that the c-number result
represents the normal ordered spectrum neglecting vacuum fluctuations. To take
these into account, we write $[{\mathcal S}_j(\omega)-1]$ in Eq.~\eqref{eq:wk}
in order to ensure that ${\mathcal S}_j(\omega)=1$ for a coherent state. The
spectrum ${\mathcal S}_j(\omega)$ can now be calculated by evaluating the
left-hand side of Eq.~\eqref{eq:wk} with the help of the corresponding matrix
elements from Eq.~\eqref{eq:cmatrix} using a two-dimensional inverse Laplace transform in $s$ and $s'$. 

As initial conditions we specify $\xi_1=0$ and real $\xi_2=\xi$, that is, the
probe field ($j=2$) is in a broad band squeezed state, whereas the pump field
is coherent. All frequency components of both pump and probe beam, are in a
squeezed/coherent vacuum ($\alpha_j=0$ for $\omega\neq\omega_{\rm L}$), only
the carrier frequencies of the two beams are displaced by real $\alpha_j$,
thus driving resonantly the atomic transitions with Rabi frequency $\Omega_j = |g_j\alpha_j|$. 
 
\begin{figure*}
\includegraphics[width=13cm]{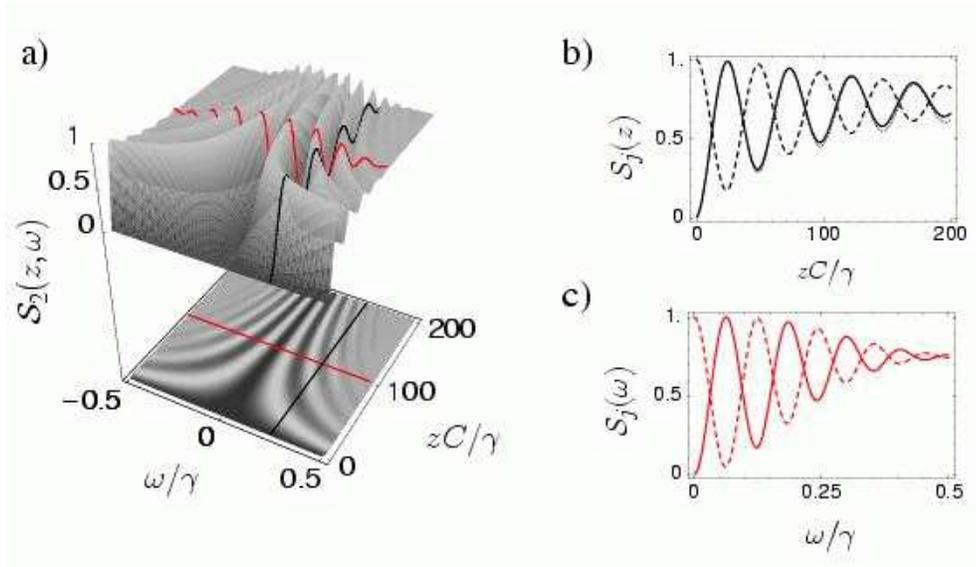}
\caption{\label{fig:mainres} Quantum properties of laser beams propagating
  through an EIT medium. (a) The fluctuation spectrum of the probe beam,
  initially (at $z=0)$ in a squeezed state with squeezing parameter $\xi=-3$
  propagates along the $z$-direction. Shown is the variance of the quadrature
  $\theta=0$ for different frequencies $\omega/\gamma$ and positions
  $z\, C/\gamma$, where C is the prefactor of Eq.~\eqref{eq:p}. (b) The
  fluctuations of pump (dashed) and probe (solid) as a function of 
  position for fixed frequency $\omega=0.25\gamma$ and $\theta=0$. For the probe, the fine dotted line shows a simulation with finite lifetime $500/\gamma$ of the ground state coherence. c)
  Fluctuation spectrum of pump and probe at fixed position $z=100\gamma/C$ and
  $\theta=0$.  (Parameters: $\Omega_1=\Omega_2=\gamma$,
  $g_1=g_2=\gamma/60$, $\xi=-3$.)} 
\end{figure*}

Using these initial conditions, we find for the fluctuation spectra of pump ($j=1$) and probe ($j=2$) 
\begin{alignat}{1}
{\mathcal S}_1(z,\omega)=&
1-\frac{f(\xi,\theta)}{\Omega^4} \Omega _1^2 \Omega _2^2\Big\{1+e^{-\gamma  P(\omega,0) z}\nonumber\\
\quad &-2 e^{-\gamma P(\omega,0)  z/2 } \cos \left[P(\omega,\Omega)\omega
  z\right]\Big\}\, ,
\label{eq:y1}\\
{\mathcal S}_2(z,\omega)=&
1-\frac{f(\xi,\theta)}{\Omega^4}\Big\{\Omega _2^4+\Omega _1^4 e^{-\gamma P(\omega,0) z}\nonumber\\
&+2 \Omega _2^2 \Omega _1^2 e^{- \gamma P(\omega,0) z/2} \cos \left[P(
  \omega,\Omega)\omega z\right] \Big\}\, ,
\label{eq:y2}
\end{alignat}
with the resonance curve  
\begin{equation}
P(\omega,\Delta)=\frac{N\left(g_1^2 \Omega_2^2+g_2^2 \Omega_1^2\right)}{c\Omega^2}\frac{|\omega^2-\Delta^2|}{(\gamma/2)^2
   \omega ^2+\left(\omega^2-\Omega^2\right)^2}.
\label{eq:p} 
\end{equation}
Furthermore, we have defined $f(\xi,\theta)=1-e^{2\xi}\cos^2\theta-e^{-2\xi}\sin^2\theta$
and $\Omega^2=\Omega_1^2+\Omega_2^2$. Indeed, for $z=0$, we find ${\mathcal
  S}_1(0,\omega)=1$ and ${\mathcal S}_2(0,\omega)=e^{2\xi}$ for the $\theta=0$
  quadrature. In Fig.~\ref{fig:mainres} (a) we have plotted the fluctuation
  spectrum of the probe pulse as a function of frequency and propagation
  length $z$. In parts (b) and (c) of Fig.~\ref{fig:mainres} cuts along the
  $\omega$ and $z$ direction are shown for both, pump and probe fields. 

We start the discussion of Eqs.~\eqref{eq:y1} and \eqref{eq:y2} with analyzing their asymptotic behavior. For $z\rightarrow\infty$ we obtain
\begin{alignat*}{1}
{\mathcal S}_1(z,\omega)\approx&1-f(\xi,\theta)\frac{\Omega_1^2 \Omega_2^2}{\Omega^4},\\
{\mathcal S}_2(z,\omega)\approx&1-f(\xi,\theta)\frac{\Omega_2^4}{\Omega^4}.
\end{alignat*}

The distance where this asymptotic behavior
is achieved is governed by the exponentials in Eqs.~(\ref{eq:y1}) and ~(\ref{eq:y2}), and is of the
order of $z_{\rm abs}=1/ \gamma P(\omega,0)$. When $\Omega_1\neq 0$, the
asymptotic quadrature fluctuations of the probe show absorption of the initial
squeezing of the field. As $P(\omega,0)$ follows qualitatively the probe field
mean value transparency curve, see Fig.~\ref{fig:res}, so does the distance where
this asymptotic behavior takes place. The pump field shows also squeezed
fluctuations in this limit, but  from the analysis of different quadratures
$\theta$ it follows that the state is no longer a minimum uncertainty
state. For equal Rabi frequencies $\Omega_1=\Omega_2$, both fields have
asymptotically the same fluctuations. Similar correlations are known in cavity EIT \cite{bib:agarwal} and from the effect of pulse-matching \cite{bib:harris,bib:fleisch}. 


In the case of strong asymmetric driving $\Omega_2\ll\Omega_1$, it follows from 
Eqs.~(\ref{eq:y1}) and ~(\ref{eq:y2}) that the pump field always stays in a
coherent state while the initial fluctuations of the probe field are
exponentially damped during propagation through the medium. The intensity of the
absorption is given by $\gamma P(\omega,0)$. As already mentioned,
$P(\omega,0)$ behaves like the mean value transparency curve, so that
we find a comparable behavior of the absorption of the probe's quadrature
fluctuations as for its mean value absorption. Thus, in this limit, our result
reproduces the analysis reported in Ref.~\cite{rv:dantan4}.

When $\Omega_1\approx \Omega_2$, however we find an entirely novel behavior. In order to bring out this new aspect most clearly, we assume $\gamma P(\omega,0) z\ll 1$, that is, we consider positions $z$ where the exponential absorption of the quadrature fluctuations can be neglected. Then, Eqs.~\eqref{eq:y1} and ~\eqref{eq:y2} can be approximated as
\begin{alignat}{1}
{\mathcal S}_{1}(\zeta,\omega)&\approx 1-f(\xi,\theta)\frac{4\Omega _1^2 \Omega _2^2}{\Omega^4} \sin^2 \zeta
\label{eq:y1ap}\\
{\mathcal S}_{2}(\zeta,\omega)&\approx 1-\frac{f(\xi,\theta)}{\Omega^4}\left\{\left(\Omega _1^2-\Omega _2^2\right)^2+4 \Omega
    _2^2 \Omega _1^2 \cos^2 \zeta\right\}
\label{eq:y2ap}
\end{alignat}
with $\zeta=z/z_{\rm osc}$, where we introduce the oscillatory length scale
$z_{\rm osc}=2/P(\omega,\Omega)\omega$. Equations~\eqref{eq:y1ap} and
\eqref{eq:y2ap} clearly display the oscillatory transfer of the initial quantum properties of the probe to the pump and back while traveling through the medium.
During this oscillatory behavior, the pump and probe field stay in a minimum
uncertainty state. Also the sum of the fluctuations ${\mathcal
  S}_{1}+{\mathcal S}_2$ is conserved. When $\Omega_1=\Omega_2$, this
oscillatory transfer is maximal and all the fluctuation properties oscillate
between the probe and the pump field during propagation in the medium. The
length scale of the oscillatory transfer $z_{\rm osc}$  can be much smaller
than the absorption length scale $z_{\rm abs}$. In Fig.~\ref{fig:mainres}, the
interplay of both scales, oscillatory and absorption, can be clearly observed. 

The oscillatory behavior implies that the outgoing quantum field can be
completely different from the incoming field, although the mean values
stay exactly the same. In other words, {\it the medium is not
transparent for the quantum properties of the field}, at least not in the sense
that the quantum state can traverse the medium unaltered. The oscillatory
behavior is qualitatively different from the absorption and is ``coherent'' since it does not imply loss of quantum properties. 

To check if these oscillations still persist with a damping rate $\Gamma_{12}$
of the ground state coherence we performed a numerical simulations. For a
typical value $\Gamma_{12}=1/500\gamma$ the resulting curve (dotted) shown in Fig.~\ref{fig:mainres} b) almost coincide with the ideal case. A complete discussion of the effect of decoherence will be published elsewhere.    

Present technology allows for an experimental observation of this effect.
The initial state can be constructed by mixing a laser tuned to the
probe optical transition with a wide band squeezed field which covers the
transparency window. With a Rabi frequency
$\Omega_1=\Omega_2=\gamma$ and an observation frequency $\omega\approx
0.1\gamma$, $z_{\rm abs}$ is ten times $z_{\rm osc}$. For a density $\rho\approx
10^9$, we have a maximum transfer of quantum properties between the fields
for $z\approx 7$cm, while $z_{\rm abs}\approx 70 $cm. When $\omega\approx
0.25\gamma$ we have a maximum transfer of quantum properties between the fields
for $z\approx 4.5$cm  ($z C/\gamma\approx 25$ in Fig.~\ref{fig:mainres} b)) while
$z_{\rm abs}\approx 12$cm ($z C/\gamma\approx 64$).
Parameters on this order are found in various EIT experiments~\cite{rv:kozuma,rv:lukincollo,rv:rmpfleisch}. 

In conclusion, we have shown that in the propagation of pump and probe beams through an EIT
medium, the quantum states are not conserved, except for the carrier
frequencies which drive the atoms on two-photon resonance. Apart from
absorption of quantum fluctuations, which takes place in approximately the same
frequency range as the absorption of the mean values, we found a novel
characteristic behavior, which is most strongly pronounced if the two beams
have comparable Rabi frequencies, consisting of an oscillatory transfer of the
initial quantum properties between the probe and the pump field. This effect
could be observed in current state-of-the-art experiments.

\acknowledgments
We thank P. Nussenzveig and P. Valente for stimulating and helpful
discussions. MB gratefully acknowledges support from the Alexander-von-Humboldt foundation.


\begin{thebibliography}{12}
\bibitem{rv:harris}
S. Harris, Phys. Today {\bf 50}, 36 (1997).
\bibitem{rv:marangos}
J. P. Marangos, J. Mod. Opt. {\bf 45}, 471 (1998).
\bibitem{rv:memoria2}
M. Fleischhauer and M. D. Lukin, Phys. Rev. A {\bf 65}, 022314 (2002).
\bibitem{rv:dantan4}
A. Dantan, A. Bramati, and M. Pinard, Phys. Rev. A {\bf 71}, 043801 (2005).
\bibitem{rv:kozuma}
D. Akamatsu, K. Akiba, and M. Kozuma, Phys. Rev. Lett. {\bf 92}, 203602 (2004).
\bibitem{bib:fleisch95} 
M. Fleischhauer and T. Richter, Phys. Rev. A {\bf 51}, 2430 (1995).
\bibitem{bib:louisell}
W. H. Louisell, Quantum statistical properties of radiation, Wiley, New York (1973).
\bibitem{bib:pablonicim}
P. Barberis-Blostein, N. Zagury,  Phys. Rev. A {\bf 70}, 053827 (2004).
\bibitem{bib:davidov}
L. Davidovich, Rev. Mod. Phys. {\bf 68}, 127 (1996).
\bibitem{bib:stochasticmethods} 
C. W. Gardiner, Handbook of Stochastic Methods, Springer-Verlag, (1994).
\bibitem{bib:agarwal}
G. S. Agarwal, Phys. Rev. Lett. {\bf 71}, 1351 (1993).
\bibitem{bib:harris}
S. E. Harris, Phys. Rev. Lett. {\bf 70}, 552 (1993). 
\bibitem{bib:fleisch}
M. Fleischhauer, Phys. Rev. Lett. {\bf 72}, 989 (1994).
\bibitem{rv:lukincollo}
M. D. Lukin, Rev. Mod. Phys. {\bf 75}, 457 (2003).
\bibitem{rv:rmpfleisch}
M. Fleischhauer, A. Imamoglu, J. P. Marangos, Rev. Mod. Phys. {\bf 77}, 633 (2005).
\end{thebibliography}
\end{document}